\documentclass[twocolumn]{aastex62}
\usepackage[super]{nth}
\usepackage[hang,flushmargin]{footmisc}

\newcommand{\logg}{\ifmmode {\log{g} }\else$\log{g}$\fi}
\newcommand{\Teff}{\ifmmode {T_{\rm eff} }\else$T_{\rm eff}$\fi}
\newcommand{\Msun}{\ifmmode {\mathrm{M}_{\odot}}\else${{M}_{\odot}}$\fi}
\newcommand{\Rsun}{\ifmmode {\mathrm{R}_{\odot}}\else${{R}_{\odot}}$\fi}
\newcommand{\Lsun}{\ifmmode {\mathrm{L}_{\odot}}\else${{L}_{\odot}}$\fi}
\newcommand{\Porb}{\ifmmode {P_{\rm orb}}\else${P_{\rm orb}}$\fi}

\newcommand{\UVA}{Department of Astronomy, University of Virginia, Charlottesville, VA 22904, USA}
\newcommand{\Vanderbilt}{Department of Physics and Astronomy, Vanderbilt University, Nashville, TN 37235, USA}
\newcommand{\NKU}{Department of Physics, Geology, and Engineering Technology, Northern Kentucky University, Highland Heights, KY 41099, USA}
\newcommand{\Salisbury}{Department of Physics, Salisbury University, Salisbury, MD 21801, USA}
\newcommand{\IAC}{Instituto de Astrof\'isica de Canarias, E-38205 La Laguna, Tenerife, Spain}
\newcommand{\ULL}{Universidad de La Laguna, Dpto. Astrof\'isica, E-38206 La Laguna, Tenerife, Spain}
\newcommand{\OCA}{Universit\'e C\^ote d'Azur, Observatoire de la C\^ote d'Azur, CNRS, Laboratoire Lagrange, France}
\newcommand{\UPitt}{Department of Physics and Astronomy and Pittsburgh Particle Physics, Astrophysics and Cosmology Center (PITT PACC), University of Pittsburgh, Pittsburgh, PA 15260, USA}
\newcommand{\Steward}{Steward Observatory, Department of Astronomy, University of Arizona, Tucson, AZ 85721, USA}
\newcommand{\Brasil}{Observat\'orio Nacional, Rio de Janeiro, Brazil}
\newcommand{\MSU}{Department of Physics, Montana State University, Bozeman, MT 59717, USA}
\newcommand{\Carnegie}{Observatories of the Carnegie Institution for Science, Pasadena, CA 91101, USA}
\newcommand{\NSFOIR}{NSF OIR Lab, Tucson, AZ 85719, USA}

\graphicspath{{./}{figures/}}

\received{July 1, 2020}
\revised{August 7, 2020}
\accepted{August 8, 2020}
\submitjournal{ApJL}

\shorttitle{Geometry of Draco~C1}
\shortauthors{Lewis et al.}

\begin{document}

\title{Geometry of the Draco~C1 Symbiotic Binary}


\correspondingauthor{Hannah M.\ Lewis}
\email{hml8vf@virginia.edu}

\author[0000-0002-7871-085X]{Hannah M.\ Lewis}
\affil{\UVA}

\author[0000-0001-5261-4336]{Borja Anguiano}
\affil{\UVA}

\author[0000-0002-3481-9052]{Keivan G.\ Stassun}
\affil{\Vanderbilt}

\author[0000-0003-2025-3147]{Steven R.\ Majewski}
\affil{\UVA}

\author[0000-0001-5611-1349]{Phil Arras}
\affil{\UVA}

\author[0000-0003-0167-0981]{Craig L.\ Sarazin}
\affil{\UVA}

\author{Zhi-Yun Li}
\affil{\UVA}

\author[0000-0002-3657-0705]{Nathan De Lee}
\affil{\NKU}
\affil{\Vanderbilt}

\author[0000-0003-3248-3097]{Nicholas W.\ Troup}
\affiliation{\Salisbury}

\author{Carlos Allende Prieto}
\affiliation{\IAC}
\affiliation{\ULL}

\author[0000-0003-3494-343X]{Carles Badenes}
\affil{\UPitt}

\author{Katia Cunha}
\affiliation{\Steward}
\affiliation{\Brasil}

\author[0000-0002-1693-2721]{D.\ A.\ Garc\'ia-Hern\'andez}
\affiliation{\IAC}
\affiliation{\ULL}

\author{David L.\ Nidever}
\affiliation{\MSU}

\author[0000-0002-7432-8709]{Pedro A.\ Palicio}
\affiliation{\OCA}
\affiliation{\IAC}
\affiliation{\ULL}

\author{Joshua D.\ Simon}
\affiliation{\Carnegie}

\author{Verne V.\ Smith}
\affil{\NSFOIR}

\begin{abstract}

Draco~C1 is a known symbiotic binary star system composed of a carbon red giant and a hot, compact companion --- likely a white dwarf --- belonging to the Draco dwarf spheroidal galaxy. 
From near-infrared spectroscopic observations taken by the Apache Point Observatory Galactic Evolution Experiment (APOGEE-2), part of Sloan Digital Sky Survey IV, we provide updated stellar parameters for the cool, giant component, and constrain the temperature and mass of the hot, compact companion.
Prior measurements of the periodicity of the system, based on only a few epochs of radial velocity data or relatively short baseline photometric observations, were sufficient only to place lower limits on the orbital period ($P > 300$\,days).
For the first time, we report precise orbital parameters for the binary system: with 43 radial velocity measurements from APOGEE spanning an observational baseline of more than 3 yr, we definitively derive the period of the system to be $1220.0^{+3.7}_{-3.5}$\,days.
Based on the newly derived orbital period and separation of the system, together with estimates of the radius of the red giant star, we find that the hot companion must be accreting matter from the dense wind of its evolved companion.


\end{abstract}


\section{Introduction} \label{sec:intro}

Symbiotic stars are interacting binaries consisting of a giant star transferring mass onto a hot, compact companion -- typically, a white dwarf (WD). 
In these systems, the fundamental power source is steady nuclear burning of accreted matter on the surface of the WD \citep[e.g.,][]{vandenheuvel1992}, and the spectrum is due to the combined emission of the photosphere of the hot companion, the cool giant, and the photoionized wind of the giant star \citep{kenyon1984,murset1999} as evidenced by the presence of strong emission lines, particularly in the Balmer series (H$\alpha$, H$\beta$, etc.)\ and in He II and higher ionization.

Of the $\sim 30$ Galactic symbiotic binaries with derived orbital parameters, the majority have relatively close orbits, with semi-major axes smaller than $\sim 0.8$\,AU and periods shorter than 2000\,days \citep{mikolajewski2003}. Furthermore, symbiotic stars tend to have nearly circular orbits, with eccentricity $e \lesssim 0.1$, though significant eccentricities have been found for systems with periods longer than 1000\,days. This observation differs from the parameters inferred for other types of binaries (i.e., non-symbiotic systems) containing late-type giants, which can show eccentric orbits for systems with orbital periods $< 1000$\,days \citep[e.g.,][]{jorissen1992}.

While symbiotic stars have been identified outside of the Milky Way (though, almost exclusively in M31 or in Milky Way/M31 satellite galaxies),
no extragalactic system has the full set of Keplerian orbital parameters (period, eccentricity, semiamplitude, barycentric velocity, and separation) derived, until now.

Draco~C1, a known symbiotic star in the Draco dwarf spheroidal (dSph) 
galaxy, is composed of a red giant (RG) 
CH carbon star \citep[i.e., showing strong CH absorption in the spectrum;][]{keenan1942,aaronson1982} and a compact companion, likely a WD \citep{munari1991,munari1994}.
Draco~C1 is classified as an $\alpha$-type symbiotic star; such symbiotic stars are characterized by their supersoft X-ray spectra, with all counts falling below $\leq 0.4$\,keV \citep{murset1997}.
This system is one of very few symbiotic systems with detected supersoft X-ray emission, as this emission is typically absorbed locally by circumstellar gas \citep{munari2019}.
As a consequence, very few $\alpha$-type symbiotic stars have been studied in the X-ray \citep[e.g., Lin 358 by][]{skopal2015b}.
Therefore, Draco~C1 provides an extraordinary opportunity.
Previous studies of the X-ray emission of this system find the WD-dominated X-ray spectrum is well fitted with a blackbody of $>10^5$\,K and a bolometric luminosity $\gtrsim 10^{38}$\,erg\,s$^{-1}$.
Together, these observations suggest stable nuclear burning on the surface of the WD \citep{munari1991,munari1994,saeedi2018}.
Further, based on data from the Optical Monitor on {XMM-Newton} \citep[OM;][]{xmmom}, \cite{saeedi2018} find long-term variability in the optical and ultraviolet (UV) emission of Draco C1, with a period $> 300$\,days. 

In this work we report stellar parameters and improved mass and radius estimates for both the primary and secondary components of the Draco~C1 system, using stellar atmosphere models fit to the spectral energy distribution (SED; Section \ref{sec:params SED}). 
In Section \ref{sec:params RVs}, we also present the first precise Keplerian orbital parameters for this system based on more than 40 spectroscopically derived radial velocities (RVs) 
from the second phase of the Apache Point Observatory Galactic Evolution Experiment \citep[APOGEE-2;][]{apogee} survey. 

The improved stellar parameters and better definition of the radial motions of the primary component of the Draco~C1 symbiotic binary enable a more comprehensive understanding of the accretion mechanism in this system (Section \ref{sec:discussion}).

\section{Data} \label{sec:data}

We utilize multi-epoch, high-resolution ($R \sim 22,500$) near-infrared (NIR; 1.51--1.70\,$\mu$m) 
spectra from the APOGEE spectrograph \citep{apogeeinstrument}, taken via APOGEE-2 as part of the Sloan Digital Sky Survey \citep[SDSS-IV;][]{2.5m,sdss4}. 
The visit-combined APOGEE spectra provide stellar parameters including effective temperature \Teff, surface gravity \logg, and metallicity [$\mathrm{M/H}$] for each target, whereas the visit-level spectra provide RVs at individual epochs \citep{apogeedatapipeline,apogeedr16}.
While the primary goal of the APOGEE survey is to measure the chemodynamical properties of stars across the Milky Way, to place these properties more broadly in the context of Galaxy evolution, the survey also targets confirmed and candidate members of 10 Local Group satellite galaxies \citep{apogee2targeting}, including stars previously identified as members of the Draco dSph.
At the conclusion of the APOGEE-2 survey, the faint members in most dSphs targeted will have received $\gtrsim 24$ visits, enabling the determination of precise orbits of identified binaries.

Draco~C1 was included on multiple plate designs, each receiving six or more visits to date. As a result, the RG component of the Draco~C1 system (2MASS: J17195764+5750054) has been observed 46 times over the duration of the APOGEE-2 survey, with those visits spanning $> 3$ yr (2016 April through 2019 June).
It is worth noting that the hot, compact companion is not detectable at infrared (IR) wavelengths 
above the flux of the RG primary, thus the RVs derived from the APOGEE spectra are representative of the velocity of the cool component of the binary.
These multi-epoch RVs, along with the Modified Julian Date (MJD) 
of the observation, associated velocity error, visit-signal-to-noise ratio (S/N), 
and individual visit spectra will all be reported as part of the final APOGEE data release (DR17, expected 2021 July); a subset of these observations are reported in the APOGEE DR16 allVisit file \citep{sdssdr16,apogeedr16}. The MJDs, RVs, and errors utilized in this work are reported in Table \ref{tab:RVs}.

For a majority of stars in APOGEE DR16, the APOGEE Stellar Parameters and Chemical Abundance Pipeline \citep[ASPCAP;][]{aspcap} pipeline derives precise stellar parameters (\Teff, \logg, metallicity, and individual chemical abundances) from the combined spectra; however, all data in the Draco dSph were not passed through ASPCAP in DR16 because the visit-spectra combination does not perform well for very faint stars (e.g., dSph members).
For this reason, we make use of the broadband photometric measurements of Draco~C1 and its companion---spanning a broad range of wavelengths, from the {XMM-Newton} soft X-rays (at $\sim 0.2$\,keV) to the {WISE} mid-IR (W3 filter at $\sim 10$\,$\mu$m)---to construct the SED, and use the stellar parameters of the best-fit stellar atmosphere model as a starting point (along with other observables, including 2MASS $JHK$ magnitude, distance, etc.) to fit a stellar isochrone.
The multiwavelength photometric data for this system are presented as the SED in Figure \ref{fig:SED}, and, along with the methods detailed in Sections \ref{sec:params SED} and \ref{sec:params RVs}, are applied to derive accurate radii and masses for the RG and WD components of the Draco~C1 system, as well as precise orbital parameters for the binary.

\section{Derived System Parameters} \label{sec:params}

\subsection{Spectral Energy Distribution} \label{sec:params SED}

\begin{figure}[!ht]
\includegraphics[trim=100 70 75 80, clip, width=1.0\columnwidth]{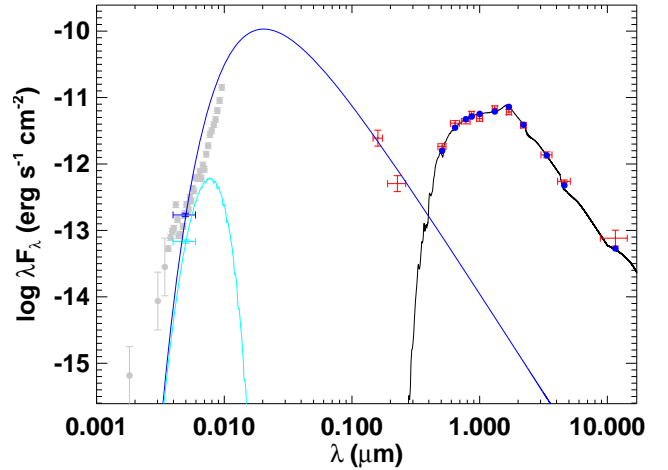}
\caption{Photometric data and associated SED fit for the symbiotic system Draco~C1. Red symbols represent observed broadband fluxes from {GALEX} FUV at 0.15\,$\mu$m to {WISE} W3 at 10\,$\mu$m. The dark blue symbol represents the {XMM} X-ray measurement in the 0.2--0.5\,keV band, corrected for extinction (the non-corrected measurement is the shown by the light blue symbol). The black curve is our Kurucz atmosphere fit to the red giant primary, and the dark blue curve is the extinction-corrected blackbody representing the white dwarf (the non-corrected version is light blue), 
scaled to match the {XMM} and {GALEX} broadband data. The full (extinction-corrected) {XMM} X-ray spectrum is shown in light gray symbols. \label{fig:SED}}
\end{figure}

Following the methods laid out by \citet{stassun2016} and \citet{stassun2017}, we fit the empirical SED of the Draco~C1 symbiotic binary.
The SED for the giant is fit with a Kurucz stellar atmosphere model \citep{kurucz2013} corrected for extinction, $A_V$.
We fit the atmosphere model to the flux measurements, minimizing $\chi^2$ by varying each parameter (\Teff, \logg, metallicity, and extinction) as well as a scaling factor -- essentially the ratio of the stellar radius to its heliocentric distance, $R_\mathrm{RG}/d$. We assume the RR Lyrae--based distance to the Draco dSph ($d = 82 \pm 2$\,kpc)
from \cite{kinemuchi2008}.
The fit to the SED (black line in Figure \ref{fig:SED}) is in good agreement with the photometric data from SDSS $g\prime$ to {WISE} W3 ($\sim 0.5-10$\,$\mu$m), with a reduced $\chi^2$ of 2.6.
From the SED fit, we find $\Teff = 3750 \pm 100$\,K, $\logg = 0.5 \pm 0.5$\,(cgs), and $\mathrm{[Fe/H]} = -1.0 \pm 0.5$, with $A_V = 0.04 \pm 0.04$.
Integrating the SED gives the bolometric flux which, together with \Teff, yields an independent empirical measure of the stellar radius of $R_{\rm RG} = 106 \pm 8$\,\Rsun. Finally, the RG radius together with the SED-derived \logg\ give an independent estimate of the stellar mass, $M_{\rm RG} = 1.1 \pm 0.6$\,\Msun.

Using the SED-derived stellar parameters, along with the 2MASS $JHK$ magnitudes, {Gaia} $G$ magnitude, and the previously mentioned distance to the Draco dSph \citep{kinemuchi2008}, we interpolate between the MESA Isochrones and Stellar Tracks \citep{dotter2016} grid of stellar isochrones using the \texttt{isochrones} package \citep{morton2015} to compute more precisely the stellar parameters, radius, and mass of the RG component of the system. The isochrone fit returns the stellar parameters listed in Table \ref{tab:params}. Most relevantly, the effective temperature, $\Teff = {3934}^{+75}_{-71}$\,K,  and bolometric luminosity, $L_\mathrm{bol, RG} = 2130^{+160}_{-400}$\,\Lsun, agree with those values estimated by \cite{akras2019} and \cite{aaronson1985}, respectively, and the isochrone fit leads to a radius of $R_\mathrm{RG} = {101.6}^{+4.7}_{-5.4}$\,\Rsun\ and a mass of $M_\mathrm{RG} = 0.735^{+0.093}_{-0.091}$\,\Msun\ for the RG. These parameters, which are consistent with those estimated from the SED analysis, are adopted for the remainder of this work.

At wavelengths shorter than SDSS $g\prime$, in particular the GALEX NUV and FUV bands and the {XMM} X-ray measurements, there is an excess flux in the SED that is contributed by the hot companion.\footnote{For completeness, in Figure \ref{fig:SED} we also show the full {XMM} spectrum; while it is broadly consistent with our simplified blackbody model, we do not include these data in our fit due to the modest departures of a true white dwarf atmosphere from a pure blackbody. By construction, the blackbody model matches the X-ray spectrum at the effective wavelength of the integrated 0.2--0.5\,keV broadband flux (dark blue symbol in Figure \ref{fig:SED}).}
Because the model atmosphere grids do not extend to wavelengths shorter than 0.1\,$\mu$m and do not extend to temperatures above $5\times10^4$\,K,
for the SED of the hot companion, we assume a simple blackbody to fit the three extinction-corrected GALEX and \textit{XMM} broadband fluxes. 
For the blackbody energy distribution, we adopt the X-ray blackbody temperature $\Teff = 1.8\times10^5$\,K and extinction column $N_H = 2.5\times10^{20}$\,cm$^{-2}$, both from \citet{saeedi2018}, 
and we assume the same distance as above for the RG.
Thus the only free parameter for the blackbody fit is a scaling factor, which corresponds to the surface area of the blackbody.
As a result, we obtain for the radius of the white dwarf 
$R_{\rm WD} = 19 \pm 6\,R_\oplus$, which agrees with the prior estimate by \cite{munari1994}.

\subsection{Radial Velocity Analysis} \label{sec:params RVs}

In order to get the most precise orbital fit for the Draco~C1 binary, we only consider the highest-quality APOGEE RV measurements.
Following several quality cuts, we are left with 43 high-quality RV measurements having associated derived uncertainties for the Draco~C1 symbiotic system. We refer to the \href{sec:appendix}{Appendix} for a detailed explanation of these constraints.
As the APOGEE visit-level RV uncertainties are known to be underestimated \citep[e.g.,][]{badenes2018}, we scale up the visit RV uncertainties, following the expression presented in C.~Brown et al.~(in prep.; see the \href{sec:appendix}{Appendix} for further details).

\begin{figure}[ht!]
\includegraphics[width=1.0\columnwidth]{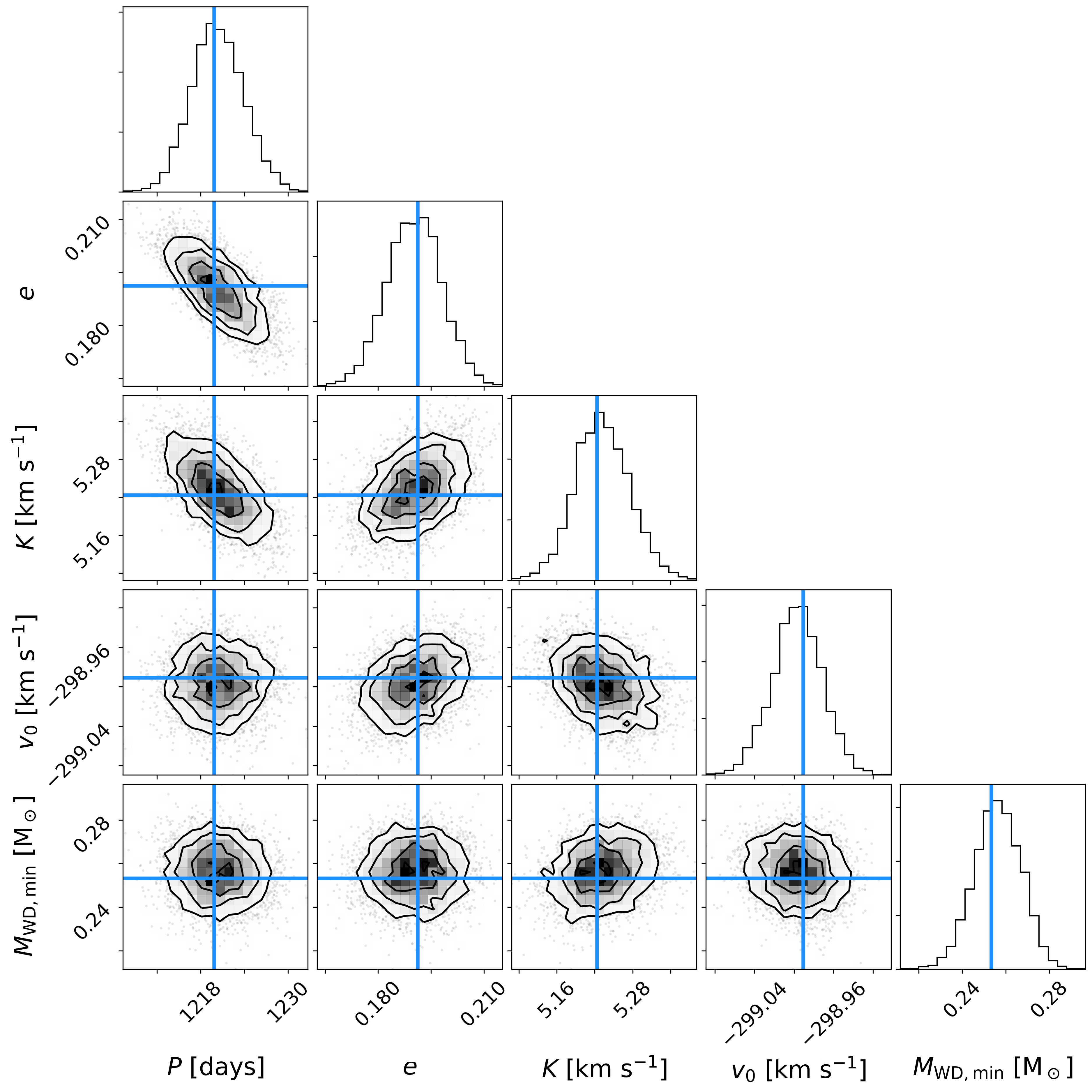}
\caption{Projections of the MCMC samples in period $P$, eccentricity $e$, semiamplitude $K$, systemic velocity $v_0$, and minimum companion mass $M_\mathrm{WD, min}$. The parameters yielded by the MAP sample are shown by the blue cross-hairs. \label{fig:corner}}
\end{figure}

\begin{figure}[ht!]
\includegraphics[width=1.0\columnwidth]{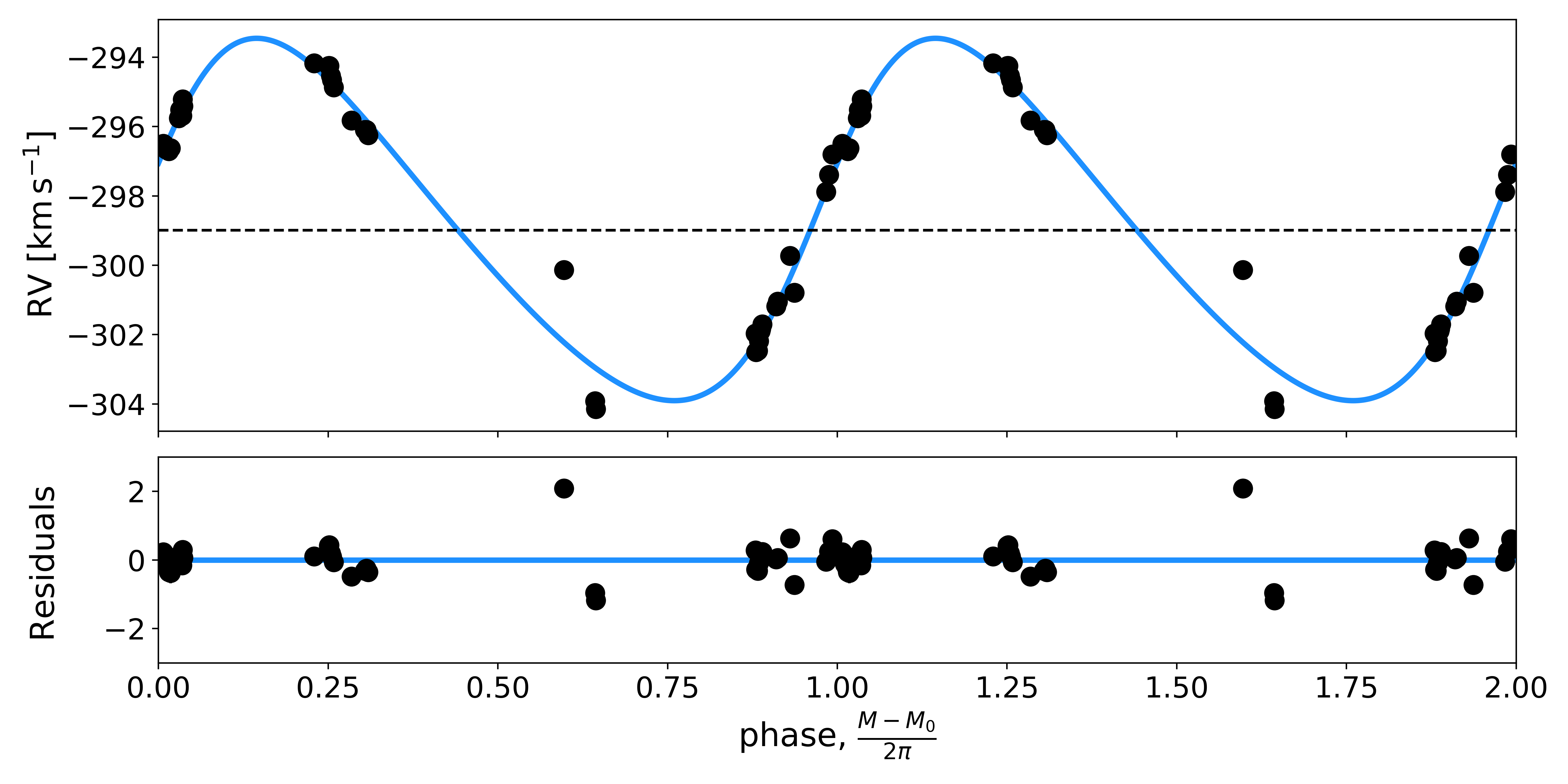}
\caption{Two full orbits of the visit velocity data from APOGEE for Draco~C1 (black points) underplotted with an orbit computed from the MAP sample returned from the MCMC analysis (blue line). Error bars on the data are typically smaller than the marker. \label{fig:RVs}}
\end{figure}

To derive orbital parameters for the system from the APOGEE RVs, we run \textit{The Joker}, a custom Monte Carlo sampler designed to produce posterior samplings over Keplerian orbital parameters that has been tested extensively on APOGEE data \citep{thejoker,pricewhelan2020}.
We generate a cache of $2^{24}$ prior samples in the nonlinear parameters described by \cite{pricewhelan2020}, evaluate the marginal likelihood of each, and perform rejection sampling to produce a minimum of $M_\mathrm{min} = 256$ posterior samples.
Following iterative rejection sampling by \textit{The Joker}, fewer than the requested number of posterior samples, $M_\mathrm{min}$, are returned for this system.
We use the few samples returned from \textit{The Joker} to initialize Monte Carlo methods to continue generating posterior samples for the Draco~C1 system \citep[for details on these methods, see][]{pricewhelan2020}.
Projections of the Markov chain Monte Carlo (MCMC) samples are shown in Figure \ref{fig:corner}, and the maximum a posteriori (MAP) 
sample is indicated.
The APOGEE RVs, phase folded to the MAP sample period, are shown in Figure \ref{fig:RVs}, and are presented in Table \ref{tab:RVs}. Note, the APOGEE RVs nearly cover a full orbital period.
The MAP sample parameters are reported in Table \ref{tab:params}, with errors given by the standard deviation of the MCMC samples.
The few RVs at phase $\sim 0.6$ that do not agree with the modeled Keplerian orbit are potentially representative of additional RV variability due to a flare or pulsation of the RG component \citep[e.g.,][]{hinkle2019}; however, there is no existing well-cadenced (i.e., observations every few days), infrared photometry that overlaps this phase in the binary orbit to confirm the occurrence of a flare.

The APOGEE RVs are indicative of a $P \sim 3.3$\,yr, noncircular ($e \sim 0.2$) orbit, with a large velocity semiamplitude $K > 5$\,km\,s$^{-1}$. 
The system barycenter velocity $v_0 = -299$\,km\,s$^{-1}$ falls within the expected dispersion about the systemic heliocentric velocity of the Draco dSph: $v_\odot = -291.0 \pm 0.1$\,km\,s$^{-1}$ \citep{walker2007} and $\sigma_v = 9.1 \pm 2.1$\,km\,s$^{-1}$ \citep{wilkinson2004}.

From the RG star mass calculated in this work and the MCMC samples, we compute the minimum WD mass, $M_\mathrm{WD, min}$, by sampling over the uncertainty on the RG mass (assuming a Gaussian noise distribution).
We find a minimum WD mass of $M_\mathrm{WD, min} = {0.253}^{+0.012}_{-0.011}$\,\Msun, which agrees with the prior mass estimate by \cite{saeedi2018}, $0.56 \pm 0.60$\,\Msun.
The WD mass function is strongly peaked around 0.56\,\Msun\ \citep{vennes1997}, implying that this system may be seen close to face-on (as $M_\mathrm{WD} = 0.56\,\Msun = 0.253\,\Msun/\sin{i}$ gives an inclination angle $i < 30^\circ$).
Further, all low-mass WDs ($M_\mathrm{WD} < 0.45\,\Msun$) are believed to be the result of enhanced mass loss in close interacting binary systems, with orbital periods on the order of a few days \citep[e.g.,][]{brown2011,rebassamansergas2011}; with a period $> 1000$\,days, it is unlikely that the WD in the Draco~C1 symbiotic system formed via this mechanism. This lends additional evidence to the Draco~C1 system being observed nearly face-on, such that $M_\mathrm{WD} \geq 0.45\,\Msun$.
The projections of the minimum companion mass versus the nonlinear parameters are shown in Figure \ref{fig:corner}; the parameters are reported in Table \ref{tab:params}.

\begin{table*}
\centering
\caption{Parameters of the Draco~C1 system. \label{tab:params}}
\begin{tabular}{llll}
\hline
\hline
Parameter & Value & Units & Reference \\
\hline

\multicolumn{3}{l}{Red Giant Parameters} \\
\hspace{3mm}\Teff & ${3934}^{+75}_{-71}$ & K & This work \\
\hspace{3mm}\logg & ${0.319}^{+0.045}_{-0.040}$ & cgs & This work \\
\hspace{3mm}[M/H] & ${-1.34}^{+0.18}_{-0.20}$ & dex & This work \\
\hspace{3mm} $L_\mathrm{bol, RG}$ & ${2130}^{+160}_{-400}$ & \Lsun & This work \\
\hspace{3mm}$R_\mathrm{RG}$ & ${101.6}^{+4.7}_{-5.4}$ & \Rsun & This work \\
\hspace{3mm}$M_\mathrm{RG}$ & ${0.735}^{+0.093}_{-0.091}$ & \Msun & This work \\

\multicolumn{3}{l}{White Dwarf Parameters} \\
\hspace{3mm}\Teff & $(1.9 \pm 0.3) \times 10^5$ & K & \cite{saeedi2018} \\
\hspace{3mm}$R_\mathrm{WD}$ & ${19}\pm{6}$ & R$_\oplus$ & This work \\
\hspace{3mm}$M_\mathrm{WD, min}$ & ${0.253}^{+0.012}_{-0.011}$ & \Msun & This work \\

\multicolumn{3}{l}{System Parameters} \\
\hspace{3mm}$A_V$ & $0.04\pm0.04$ & mag & This work \\
\hspace{3mm}d & ${82}\pm{2}$ & kpc & \cite{kinemuchi2008} \\
\hspace{3mm}$P$ & $1220.0^{+3.7}_{-3.5}$ & days & This work \\
\hspace{3mm}$e$ & $0.1906^{+0.0076}_{-0.0078}$ & & This work  \\
\hspace{3mm}$K$ & $5.224^{+0.045}_{-0.041}$ & km\,s$^{-1}$ & This work \\
\hspace{3mm}$v_0$ & $-298.991^{+0.025}_{-0.026}$ & km\,s$^{-1}$ & This work \\
\hspace{3mm}$a_\mathrm{min}$ & $2.227^{+0.071}_{-0.069}$ & AU & This work \\
\hline
\end{tabular}
\end{table*}

\section{Discussion} \label{sec:discussion}

This work provides the first detailed study of the orbital parameters of the Draco~C1 symbiotic binary, as well as the most precise constraints on the stellar parameters to date---including temperature, mass, and radius---for the cool RG and hot WD components of the system.
Of all confirmed extragalactic symbiotic stars (to date, $\sim 75$ systems), $< 25$\% of systems have prior constraints placed on the orbital period, only $\sim 5$\% have estimated masses for the hot compact companions (including WDs and neutron stars), and no other extragalactic system has a precisely derived Keplerian orbit \citep{merc2019}.

The 1220 day period derived in this work for the Draco C1 binary
falls into the typical range for observed Milky Way symbiotic systems ($P < 2000$\,days); however, the orbit is non-circular---though this is to be expected for systems with periods longer than 1000\,days---and has a significantly wider minimum separation than similar (in eccentricity-period space) Galactic symbiotic systems. 
Because symbiotic stars have the largest orbital separations of all interacting binaries, their study is relevant to understanding the 
early evolution of detached (e.g., double degenerate systems)
and semidetached (e.g., cataclysmic variables) binary stars.

\begin{figure}[ht!]
\includegraphics[width=1.0\columnwidth]{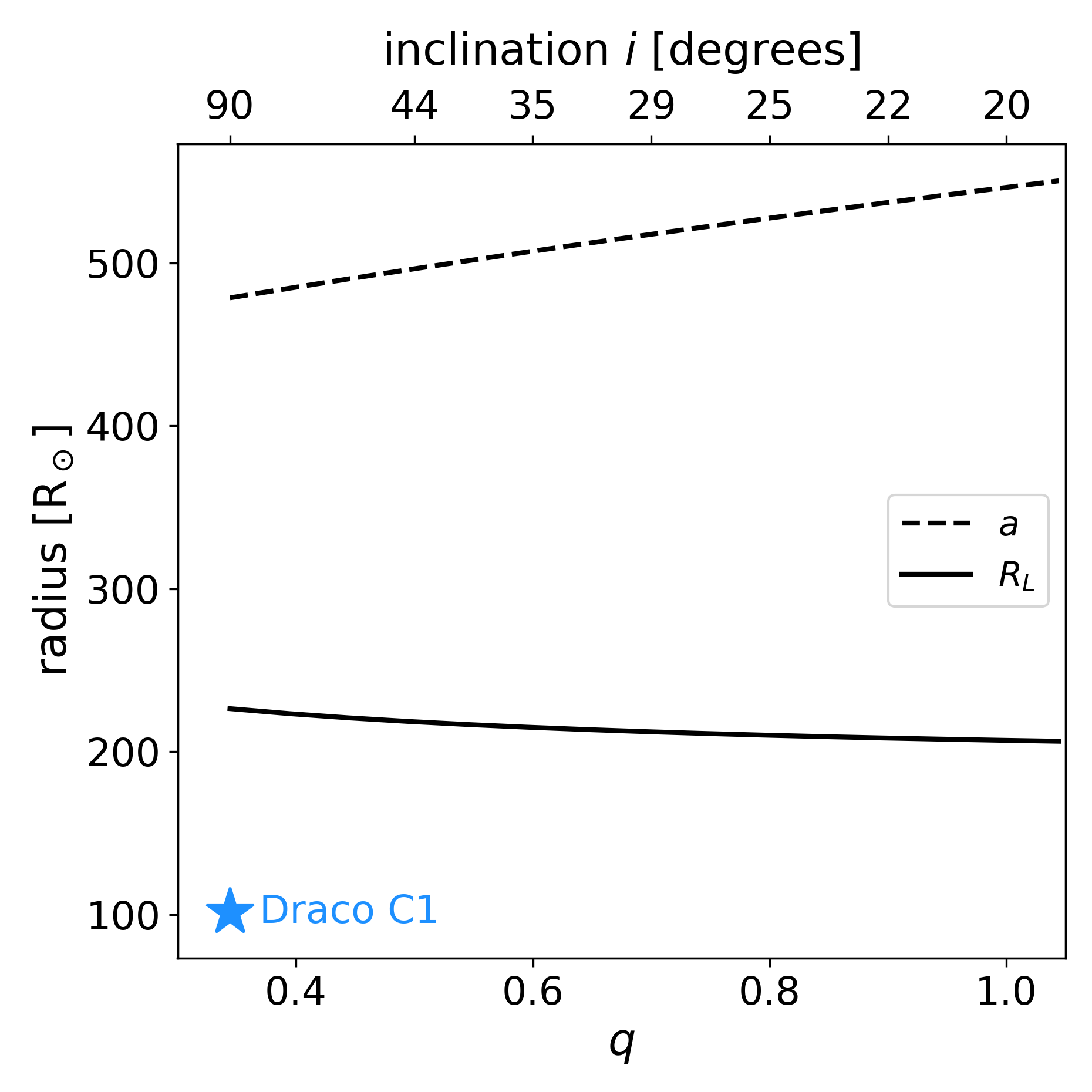}
\caption{Radius of the Roche lobe, $R_L$, of the RG (solid line) and the semi-major axis, $a$, of the Draco~C1 orbit (dashed line) versus the mass ratio, $q$ (lower $x$-axis), and inclination, $i$ (upper $x$-axis). The radius of the photosphere of the Draco~C1 RG $R_\mathrm{RG}$ and minimum mass ratio $q = 0.344$ are indicated by the blue star.
Note that $R_L$ depends very weakly on $q$ (and therefore $i$), so that for any inclination
the orbital separation is expected
to be large enough that the RG's photosphere is inside the Roche lobe. \label{fig:RLOF}}
\end{figure}

Applying Kepler's third law,
\begin{equation}
a = \left( \frac{G M_\mathrm{RG} P^2 (1 + q)}{4 \pi^2} \right)^{1/3},
\end{equation}
where the mass ratio $q = M_\mathrm{WD}/M_\mathrm{RG}$ and $M_\mathrm{WD, min} = M_\mathrm{WD} \sin{i}$, we calculate the binary separation $a$, as a function of mass ratio. Furthermore, we can calculate the radius $R_L$ of the RG star required for Roche lobe overflow (RLOF) following \cite{eggleton1983}, such that 
\begin{equation}
R_L = \frac{0.49 q^{-2/3}}{0.6 q^{-2/3} + \ln{(1 + q^{-1/3})}} \times a \label{eq:RLOF radius},
\end{equation}
which assumes a circular orbit.
If we assume an inclination of $i = 90^\circ$ ($\sin{i} = 1.0$), 
then $M_\mathrm{WD, min} = M_\mathrm{WD} = 0.253\,\Msun$ and $q = 0.253\,\Msun/0.735\,\Msun = 0.344$;
this leads to a minimum orbital separation of $a = 2.227$\,AU (reported in Table \ref{tab:params}) and the Roche lobe radius of the RG star $R_L \sim 225\,\Rsun$ (1.05\,AU).
Calculating the orbital separation $a$ and Roche lobe radius $R_L$ of the RG star for a range of inclinations $i$ down to $i \sim 20^\circ$ (i.e., up to $q = 1.0$),
we show (Figure \ref{fig:RLOF}) that, for the RG properties and orbital parameters derived in this work, the orbital separation of the binary is large enough that the photosphere of the RG is well inside its Roche lobe.

Based on Figure \ref{fig:RLOF}, for mass ratios up to $q = 1.0$ the Roche lobe radius $R_L \sim 2 \times R_\mathrm{RG}$. For this reason, it is unlikely that the accretion onto the WD is due to standard RLOF; this conclusion conflicts with the mass-transfer model suggested by \cite{saeedi2018}.

We suggest that the system is undergoing wind Roche lobe overflow (WRLOF), where the dense stellar wind from the RG companion is filling the Roche lobe, instead of the star itself \citep{mohamed2007,mohamed2012}.
Accretion onto the WD is enhanced by focusing of the stellar wind from the RG toward the binary orbital plane, and the WD can accrete at rates of up to $10^{-7}$\,\Msun\,yr$^{-1}$, and thus power its high luminosity \citep{skopal2015}.
The WD accretion rate calculated by \cite{munari1994} to explain the luminosity of the hot component (assuming stable H-burning of the accreted material) in the Draco~C1 symbiotic system is $\lesssim 10^{-7}$\,\Msun\,yr$^{-1}$, so accretion via WRLOF is the most likely mass-transfer scenario taking place there.

\section{Summary} \label{sec:summary}

Symbiotic stars offer an ideal astrophysical laboratory for detailed studies of wind accretion and mass transfer, as the large temperature gradient between the two binary components allows observations of accretion processes over a broad range of wavelengths \citep[e.g.,][]{skopal2015b}.
As the only extragalactic symbiotic binary with precise orbital parameters and stellar parameters available, the Draco~C1 system provides a unique testbed for future models of wind-mass transfer.

The key results of this work are summarized below:
\begin{enumerate}
    \item Based on the fit to the SED and stellar isochrone, the carbon RG component 
    of the Draco~C1 symbiotic has a radius of $\sim 100$\,\Rsun\ and a mass of $\sim 0.7$\,\Msun.
    \item The orbital period of the system, $1220.0^{+3.7}_{-3.5}$\,days, places the RG star well within its Roche lobe radius, indicating that mass transfer most likely follows a wind-accretion model for symbiotic binaries like that presented by \cite{skopal2015}.
    \item To date, the Draco~C1 symbiotic binary is the only extragalactic symbiotic star system with precisely derived Keplerian orbital parameters \citep{merc2019}, and contributes to the $< 5$\% of {\it all}
    systems with precise limits placed on the mass of the WD secondary.
\end{enumerate}

\acknowledgments

We thank the referee for the detailed review of this Letter and constructive comments that have helped improve the presentation of the results.
We also thank Adrian Price-Whelan for helpful discussions and suggestions.
This research made use of the New Online Database of Symbiotic Variables \citep{merc2019} and TOPCAT \citep{topcat}.

H.M.L., B.A., and S.R.M. acknowledge support from National Science Foundation grant AST-1616636.
B.A. also acknowledges the AAS Chrétien International Research Grant.
Z.Y.L. is supported in part by NASA 80NSSC20K0533.
N.D. would like to acknowledge support by the National Science Foundation grant AST-1616684.
C.B. acknowledges support from National Science Foundation grant AST-1909022.
D.A.G.H. acknowledges support from the State Research Agency (AEI) of the Spanish Ministry of Science, Innovation and Universities (MCIU) and the European Regional Development Fund (FEDER) under grant AYA2017-88254-P.
P.A.P. acknowledges support from the ANR 14-CE33-014-01 project.

Funding for the Sloan Digital Sky Survey IV has been provided by the Alfred P. Sloan Foundation, the U.S. Department of Energy Office of Science, and the Participating Institutions. SDSS acknowledges support and resources from the Center for High-Performance Computing at the University of Utah. The SDSS website is \url{www.sdss.org}.

SDSS is managed by the Astrophysical Research Consortium for the Participating Institutions of the SDSS Collaboration including the Brazilian Participation Group, the Carnegie Institution for Science, Carnegie Mellon University, the Chilean Participation Group, the French Participation Group, Harvard-Smithsonian Center for Astrophysics, Instituto de Astrof\'isica de Canarias, The Johns Hopkins University, Kavli Institute for the Physics and Mathematics of the Universe (IPMU) / University of Tokyo, the Korean Participation Group, Lawrence Berkeley National Laboratory, Leibniz Institut f\"ur Astrophysik Potsdam (AIP), Max-Planck-Institut f\"ur Astronomie (MPIA Heidelberg), Max-Planck-Institut f\"ur Astrophysik (MPA Garching), Max-Planck-Institut f\"ur Extraterrestrische Physik (MPE), National Astronomical Observatories of China, New Mexico State University, New York University, University of Notre Dame, Observat\'orio Nacional / MCTI, The Ohio State University, Pennsylvania State University, Shanghai Astronomical Observatory, United Kingdom Participation Group, Universidad Nacional Autónoma de M\'exico, University of Arizona, University of Colorado Boulder, University of Oxford, University of Portsmouth, University of Utah, University of Virginia, University of Washington, University of Wisconsin, Vanderbilt University, and Yale University.

%

\vspace{5mm}


\software{
          isochrones \citep{morton2015},
          thejoker \citep{thejoker,pricewhelan2019,pricewhelan2020},
          TOPCAT \citep{topcat}
          }



\appendix
\setcounter{table}{0}
\renewcommand{\thetable}{A\arabic{table}}

\section*{APOGEE Flags and RV Visit Uncertainties} \label{sec:appendix}

Here we detail the APOGEE flags used to remove low-quality data from our sample.
First, we require that the visit-level (allVisit) \texttt{STARFLAG} does not contain the \texttt{LOW\_SNR} flag (bitmask value: 4), such that only visits with  $\mathrm{S/N} > 5$ contribute to the fit. Additionally, we require that the \texttt{STARFLAG} bitmask does not contain \texttt{VERY\_BRIGHT\_NEIGHBOR}, \texttt{PERSIST\_HIGH}, \texttt{PERSIST\_JUMP\_POS}, or \texttt{PERSIST\_JUMP\_NEG} (bitmask values: 3, 9, 12, 13). These bitmasks remove the most obvious data reduction or calibration failures due to otherwise poor data.

Since APOGEE visit-level RV uncertainties (\texttt{VRELERR} in the allVisit file) are known to be underestimated \citep[e.g.,][]{badenes2018}, 
we consider the expression
\begin{equation}
\sigma_\mathrm{RV}^2 = (3.5 (\texttt{VRELERR})^{1.2})^2 + (0.072\,\mathrm{km\,s}^{-1})^2
\end{equation}
presented in C.~Brown et al.~(in prep.), where $\sigma_\mathrm{RV}$ is the inflated visit velocity error for a given visit, to impose a minimum error of $0.072$\,km\,s$^{-1}$. The 43 APOGEE RVs meeting the quality cuts described above, and the associated errors $\sigma_\mathrm{RV}$, are reported in Table \ref{tab:RVs}.

\begin{table*}[h]
\centering
\caption{RV measurements from APOGEE. \label{tab:RVs}}
\begin{tabular}{ccc}
\hline
\hline
MJD & RV & $\sigma_\mathrm{RV}$\\
 & (km s$^{-1}$) & (km s$^{-1}$) \\
\hline

57495.5011 & -297.886 & 0.114  \\
57500.4763 & -297.394 & 0.121  \\
57506.4454 & -296.802 & 0.143  \\
57524.3284 & -296.498 & 0.110  \\
57527.3767 & -296.579 & 0.093 \\
\vdots & \vdots & \vdots \\
\hline
\end{tabular}
\tablecomments{This table is available in its entirety in machine-readable form.}
\end{table*}

\bibliographystyle{aasjournal}
\bibliography{main}

\end{document}